\documentclass[12pt]{iopart}

\usepackage[utf8]{inputenc}
\usepackage{cancel}
\usepackage[left=0.75in,right=0.75in,top=0.75in,bottom=0.75in]{geometry}
\usepackage{setspace}
\usepackage{graphics}
\usepackage{amssymb}
\usepackage{amsmath}
\usepackage{bm}
\usepackage{mathtools}
\usepackage{times}
\usepackage{color}
\usepackage{cite}
\usepackage{graphicx}

\newcommand{\dee}{{\rm d}}
\newcommand{\calS}{{\mathcal S}}
\newcommand{\calO}{{\mathcal O}}

\newcommand{\calD}{{\mathcal D}}
\newcommand{\calJ}{{\mathcal J}}

\newcommand{\calP}{{\mathcal P}}
\newcommand{\eqa}{\begin{eqnarray}}
\newcommand{\eeqa}{\end{eqnarray}}
\newcommand{\eq}{\begin{equation}}
\newcommand{\eeq}{\end{equation}}
\newcommand{\ex}{{\rm e}}
\newcommand{\kT}{k_{\rm B}T}

\begin{document}

\title{Stochastic path power and the Laplace transform}

\author{S.P. Fitzgerald$^{1,*}$, T.J.W. Honour$^1$}

\address{                    
   $^1$Department of Applied Mathematics, University of Leeds, LS2 1JT, United Kingdom
}
\ead{$^*$S.P.Fitzgerald@leeds.ac.uk}

\begin{abstract}Transition probabilities for stochastic systems can be expressed in terms of a functional integral over paths taken by the system. Evaluating the integral by the saddle point method in the weak-noise limit leads to a remarkable mapping between dominant stochastic paths and conservative, Hamiltonian mechanics in an effective potential. The conserved ``energy'' in the effective system has dimensions of power. We show that this power, $H$, can be identified with the Laplace parameter of the time-transformed dynamics.  This identification leads to insights into the non-equilibrium behaviour of the system. Moreover, it facilitates the explicit summation over families of trajectories, which is far harder in the time domain. This is crucial for making contact with the long-time equilibrium limit.
\end{abstract}

\maketitle

\section{Introduction}
Functional- or path-integral techniques were first developed by Wiener in the 1920s as a means to study Brownian motion. Since Feynman's seminal work in the 1950s however \cite{FH}, they have been more widely known for their use in quantum mechanics and quantum field theory. They have also been used to great effect in classical statistical physics and field theory; see for example \cite{MSR,DP}. Their application to the modern study of classical stochastic processes can be traced back to the work of Onsager and Machlup \cite{onsager}, Stratonovich \cite{stratonovich}, and Graham \cite{graham}. The generalization to correlated noise was accomplished in the 1990s by McKane and co-workers \cite{mckane,mckane3}, and a general field-theory based formalism was developed by Hochberg et al \cite{hochberg}. For an introduction to path integral methods for stochastic processes, see Wio \cite{wio}, and for path integral methods in general Schulman \cite{schulman} and Kleinert \cite{kleinert}. 

For the quantum-mechanical case, in the semiclassical limit of small $\hbar$, the dominant system trajectories are small fluctuations about the classical paths. In this limit, even the wholly quantum-mechanical phenomenon of barrier penetration can be understood through classical trajectories in the inverted potential $V\to-V$, known as instantons \cite{coleman}. An analogous correspondence also exists for stochastic processes, where most-probable paths in a potential $V$ correspond to Hamiltonian trajectories in an effective potential $-|\nabla V|^2$ \cite{caroli,ge} and much of the formalism can be  carried over \cite{einchcomb}. Action-minimization methods to find the most probable path through higher-dimensional potentials have been developed \cite{gMAM,olender,koehl,grisell,kikuchi2020,fitzgerald2023}. The majority of previous work has focused on determining the infinite-time paths, and hence average transition rates. However, the full temporal information need not necessarily be discarded, and indeed is required for the calculation of quantities such as first-passage densities, and for situations where local equilibrium cannot be assumed.  

In this paper, after a brief review of the formalism, we re-examine the stochastic--deterministic correspondence, and identify the conserved quantity in the effective mechanics as the Laplace parameter in the time-transformed diffusive dynamics. We illustrate the utility of the approach with the simple examples of a particle confined to an interval in one dimension, and the harmonic oscillator.

\section{Formalism}

The starting point is an overdamped Markovian stochastic process driven by Gaussian white noise, defined by the Langevin equation 
\eq
\gamma\dot x(t) = -V'(x) + \xi(t);\;\;\langle\xi(t)\xi(t')\rangle=2D\delta(t-t')
\eeq or
\eq
\gamma dX_t = -V'(X_t)dt+\sqrt{2D}dW,
\eeq with $dW$ the increments of the standard Wiener process. $V(x)$ is the potential, $D$ the noise strength, and $\gamma$ a dissipation parameter. The fluctuation-dissipation theorem gives $D=\gamma\,\kT$. The first equation above is not valid in the limit of zero correlation time, but it suggests how to define an action based on the noise probability density functional:
\eq
\calP[\xi]\sim\exp\left( -\frac1{4D}\int_0^t \xi(t')^2\dee t'\right). \label{eq:noise}
\eeq This in turn leads to the following expression for the weight $\calP[x]$ attached to a stochastic path $x(t)$:
\eq
\calP[x]\sim\exp\left( -\frac1{4D}\int_0^t \left(\gamma\dot x+V'  \right)^2\dee t'\right)\equiv \exp\left( -\frac{\calS[x]}{4D}\right), 
\eeq defining the stochastic action $\calS$. The probability density function $\rho(x,t)$ satisfies the Smoluchowski equation 
\eq
\frac{\partial \rho}{\partial t} = \frac{\partial}{\partial x}\left(\rho V'+D\frac{\partial \rho}{\partial x}\right);\;\;\rho(x,0) = \delta(x-x_0).\label{eq:smol}
\eeq $\rho(x,t)$ corresponds to the conditional transition probability density $P(x,t|x_0,0)$. This density can be written in terms of a functional integral:
\eq
\rho(x,t) = P(x,t|x_0,0) = \int\calD x\,\calJ[x] \exp\left( -\frac{\calS[x]}{4D}\right), \label{eq:pi}
\eeq where $\calJ = \exp\left( \frac1{2\gamma}\int_0^tV''(x(t'))\dee t'\right)$ is the functional Jacobian arising from the implicit change of variables $\xi\to x$\footnote{See e.g. \cite{wio}. If we interpreted (1) as an It\^o SDE, this Jacobian would equal one. However, the cross term in the action integral $2\gamma V'\dot x\dee t = 2\gamma V'\dee x $ would then be an It\^o integral, resulting in an additional term exactly equivalent to the non-unit Jacobian as given above.}. The integral is taken over paths $x(t')$ satisfying $x(0)=x_0,x(t)=x$, and an infinite normalization constant (arising from the ``$\sim$'' in \eqref{eq:noise}
 above) has been absorbed into the measure. As $D\to 0$, the integral is dominated by paths lying close to the smooth minimizer of $\calS$, which we call $x_*(t)$. This satisfies the Euler-Lagrange equation for $\calS$:
\eq
2\gamma^2\ddot x_* = 2V'V'',\label{eq:EL}
\eeq that is, the conservative Hamiltonian dynamics of a particle of mass $2\gamma^2$ moving in the effective potential $F=-V'^2$. This dynamics has a conserved quantity
\eq
H = \gamma\dot x_*^2 - V'(x_*)^2/\gamma\longrightarrow \dot x_*^2 - V'^2,
\eeq ($2\gamma^{-1}\times$) the energy of the effective system, which has dimensions of power. We set $\gamma=1$ and call $H$ the {\it path power}. 

As $D\to 0$, we can approximate the expression for $\rho$ by restricting the integral in \eqref{eq:pi} to paths ``near'' the minimizer $x_*$. More precisely, quadratic fluctuations around $x_*$ may be integrated over by writing a general path $x(t)$ as $x(t) = x_*(t)+y(t)$ and expanding the action to second order in $y(t)$. This results in a Gaussian functional integral that can be performed explicitly, and yields the following expression for $\rho$:
\eq
\rho(x,t) = \left(4\pi D\,{\rm det}\hat L  \right)^{-1/2}\calJ[x_*]\exp\left( -\frac{\calS[x_*]}{4D}\right), \label{eq:rho}
\eeq valid as $D\to 0$. The operator $\hat L$ is the second variation of the action, and its determinant can be evaluated by various techniques; see for example \cite{schulman,tarlie}. If more than one solution for $x_*$ exists, $\rho$ will contain a term similar to the above for each of them. 

\section{$H$ and the Laplace transform} 
Inserting the minimizer $x_*$ into the action functional $\calS[x]$ gives Hamilton's principal function $S(x,t)$:
\eq
S(x,t) = 2\Delta\! V - Ht + 2\int_{p}\sqrt{H + V'^2}\,\dee s. 
\eeq The term $\Delta\! V = V(x) - V(x_0)$ is path-independent and results from the total derivative in the action. $p$ is the path, satisfying eq.\eqref{eq:EL} -- for a simple direct path in one dimension, this is simply from $x_0$ to $x_1$ along the $x$-axis \cite{ge}; for paths involving turns, or in higher dimensions, it is more complicated. $H$ is as above, and is defined implicitly via
\eq
0 = \frac{\partial S}{\partial H} = -t + \int_{p}\frac{\dee s}{\sqrt{H + V'^2}}, \label{eq:H}
\eeq as can also be seen by integrating the energy equation of the effective classical mechanics $\dot x_* = \pm\sqrt{H+V'(x_*)^2}$. The fluctuation determinant det$\,\hat L$ is given by (more details are given in Appendix A)
\eqa
{\rm det}\,\hat L = x
 & = & \sqrt{H+V'(x_0)^2}\sqrt{H+V'(x)^2}\; \int_{p}\frac{\dee s}{(H + V'^2)^{3/2}}\nonumber\\
 & = & \sqrt{H+V'(x_0)^2}\sqrt{H+V'(x)^2} \; \, 2\,\left|\frac{\partial^2 S}{\partial H^2}\right|.\label{eq:detL}
\eeqa The appearance of the derivatives of $S$ with respect to $H$ is highly suggestive. This is underlined by writing $S$ as a Legendre transform
\eq
S(x,t) = -Ht + W(x,H), \label{eq:legendre}
\eeq defining $W$ (Hamilton's characteristic function for the effective mechanics). Extracting the relevant factors from $\rho$ in \eqref{eq:rho}, we can write
\eq
\rho(x,t) \propto \exp\left( +\frac{Ht}{4D} \right)\exp\left( -\frac{W(x,H)}{4D} \right)\sqrt{\frac{8\pi D}{|W''|}},
\eeq where $H$ satisfies \eqref{eq:H}. This is precisely an inverse Laplace transform performed via steepest descents in the $D\to 0$ limit. Identifying $H/4D$ as the transform parameter, we can immediately read off
\eq
\bar \rho\left(x;\frac{H}{4D}\right) \propto \frac{\calJ\exp\left( -  \frac{W(x,H)}{4D} \right)}{\sqrt{\sqrt{H+V'(x_0)^2}\sqrt{H+V'(x)^2}}}\equiv A \exp\left( -  \frac{W(x,H)}{4D} \right), \label{eq:rhobar}
\eeq defining the prefactor $A$. $\bar\rho$ is the Laplace transform of $\rho$, and also emerges from a WKB analysis of the Laplace transform of \eqref{eq:smol} (Appendix B).

The identification of the (noise-scaled) path power as the Laplace parameter, the variable conjugate to time, leads to a number of observations. Firstly, by the final value theorem, the $H\to 0$ limit corresponds to $t\to\infty$. For an uphill path segment, where $\sqrt{V'^2} = +V'$, $W\to 4\Delta\! V$, and the familiar Kramers form $\exp\left(  -\Delta\! V/D\right)$ emerges -- the long time average rate for a process driven by noise of strength $D$ to climb a potential barrier of height $\Delta\! V$. This relies on the assumption of quasi-equilibrium being reached at the bottom of the barrier, which we identify with $H\to 0$. 
However, all the finite-time, non-equilibrium information remains when $H$ is left finite. The awkwardness of the implicit definition of $H(t)$ in \eqref{eq:H} is no longer an issue when working in the Laplace domain. This is particularly helpful when multiple trajectories' contributions need to be summed, as the following example demonstrates. 

\section{Example: free diffusion with reflecting boundaries}

\begin{figure}
\centering
\includegraphics[width=0.8\textwidth]
{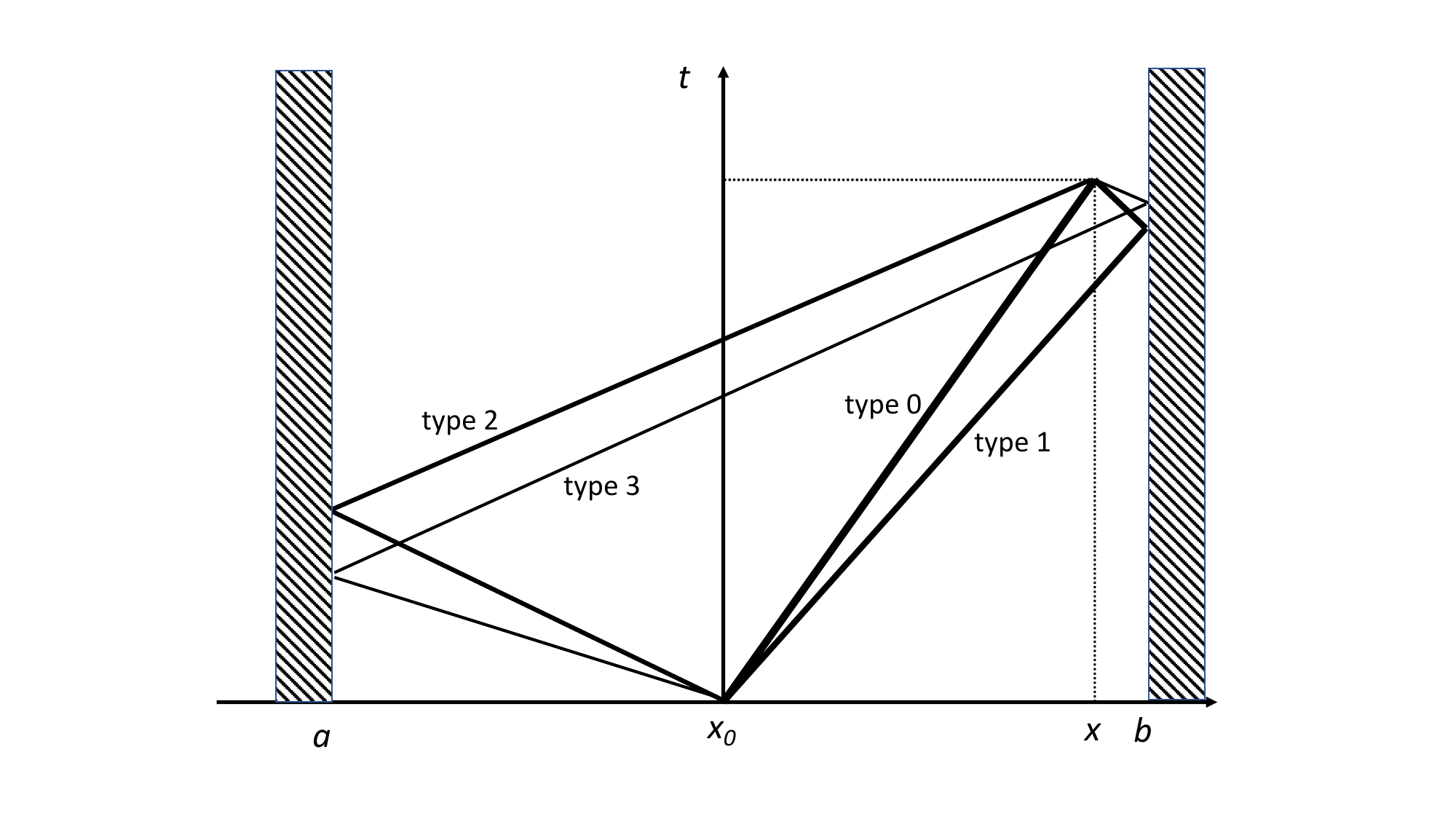}
\caption{The four ways to reach $x$ from $x_0$ with reflecting boundaries at $x=a,b$. A general path involving $n$ complete circuits of the interval $[a,b]$ must leave $x_0$ in a certain direction, and arrive at $x$ from a certain direction, giving rise to the four types of path.}
\label{fig:interval}
\end{figure}

Consider a particle diffusing on the $x$-axis, with $x=x_0\in (a,b)$ at $t=0$, and reflecting boundaries at $x=a,b$ with $b>a$. This situation is analogous to the ``particle in a box'' problem from elementary quantum mechanics. The density $\rho(x,t)$ satisfies the diffusion equation with zero-flux boundary conditions:
\eq
\frac{\partial \rho}{\partial t}= D\frac{\partial ^2\rho}{\partial x^2};\;\;\rho(x,0)=\delta(x-x_0);\;\;\frac{\partial \rho}{\partial x}(a,t) = 0 = \frac{\partial \rho}{\partial x}(b,t), \;\text{or}
\eeq
\eq
s\bar\rho - \delta(x-x_0) = D\frac{\partial ^2\bar\rho}{\partial x^2}
\eeq for the transformed density $\bar\rho(x;s)$, with Laplace parameter $s$ identified as $H/4D$. Since $V=0$ here, the effective potential is also 0, the Jacobian $\calJ = 1$, and everything can be calculated explicitly. $A = H^{-1/2}$, and $W = 2\sqrt H |p|$ where $|p|$ is the total distance travelled, i.e. $|x-x_0|$ for a simple direct path from $x_0$ to $x$. Such a path gives 
\eq
\bar\rho(x;s)  = \frac{\ex^{-2\sqrt{H}|x-x_0|/4D}}{\sqrt{H}}= \frac{\ex^{-|x-x_0|\sqrt{s/D}}}{\sqrt{4Ds}} = {\rm L.T.}\; \left( \frac{\ex^{-(x-x_0)^2/4Dt}}{\sqrt{4\pi Dt}}\right)
\eeq as expected (this can also be obtained directly from the untransformed expressions in the previous section). So, a simple direct path from $x_0$ to $x$ gives the expected solution for the infinite region $a\to -\infty,b\to\infty$. With reflecting boundaries at $x=a,b$, however, there are four Hamiltonian trajectories that reach $x$ from $x_0$: a bounce off $x=b$ (which we call ``type 1''), a bounce off $x=a$ (``type 2''), and a bounce off both (``type 3''), plus the simple direct path (``type 0''), as shown in Fig.\ref{fig:interval}. 

These have actions, numbered by type, given by 
\eqa
W_0/2\sqrt H & = & |x-x_0|\to x_0-a,\; b-x_0\;\;{\rm as}\;\;x\to 
a,b\nonumber\\
W_1/2\sqrt H & = & 2b-x_0-x\to 2b-a-x_0,b-x_0\;\;{\rm as}\;\;x\to 
a,b\nonumber\\
W_2/2\sqrt H & = & x+x_0-2a\to x_0-a,b-2a+x_0\;\;{\rm as}\;\;x\to 
a,b\nonumber\\
W_3/2\sqrt H & = & |x_0-x|+2b-2a\to 2b-a-x_0,x_0+b-2a\;\;{\rm as}\;\;x\to 
a,b.
\eeqa where we also give the limiting values of the actions at the boundaries. This leads to 
\eq
\bar\rho =  H^{-1/2}\left(\exp (-W_0/4D) + \exp (-W_1/4D) + \exp (-W_2/4D )
+ \exp( -W_3/4D) \right). 
\eeq For $x>0$, types 0 and 2 are increasing functions of $x$ while types 1 and 3 are decreasing. This simply corresponds to the two independent solutions of the second-order ODE, and introduces relative sign changes in the $x$-derivatives. By linearity, $\bar\rho$ as above satisfies the transformed diffusion equation, and is zero-flux at $x=a$ or $b$ because of the way the terms in $\partial\bar\rho/\partial x$ cancel in pairs: $W_1(a)=W_2(a); W_3(a) = W_4(a)$ and $W_1(b)=W_3(b);W_2(b)=W_4(b)$. The initial condition can be checked (without Laplace inversion) using the initial value theorem:
\eq
\lim_{s\to\infty} s\bar\rho(x,s) = \lim_{t\to 0}\rho(x,t).
\eeq Because $W_0<W_{1,2,3}$ at fixed $H$, as $s$ (or $H$) goes to infinity, only the first term contributes to the limit, recovering the infinite region solution, and hence the short-time limit is correctly captured. However, the long-time limit is incorrect: the final value theorem
\eq
\lim_{s\to 0} s\bar\rho(x,s) = \lim_{t\to \infty}\rho(x,t)
\eeq gives 
\eq
\frac{H}{4D}\,\bar\rho\sim\frac{\sqrt{H}}{D}\; \text{as}\; H\to 0;\; \rho(x,t\to\infty)\to 0,
\eeq 
as in the infinite region case, whereas the reflecting boundaries should give a uniform distribution in this limit: $\rho(x,t\to\infty) = 1/(b-a)$.

This apparent contradiction can be resolved by including the contributions of more paths. For each of 
the types of path above, we can add $n$ whole circuits from $x_0$ to 
$x_0$. Each of these will accrue additional action $ n\times 2\sqrt H\times 2(b-a)$. In the transformed domain, we work with a fixed $H$, so all these paths have the same $H$ (in contrast to them all having the same $t$ in the time domain). This means that the exponentials form a 
geometric series, and the multi-bounce paths can be summed over explicitly, resulting in 
\eqa
\bar \rho & = & \frac{\exp( -W_0/4D) + \exp( -W_1/4D) + \exp( -W_2/4D )
+ \exp( -W_3/4D )}{\sqrt H\left(1 - \exp -\sqrt H 
(b-a)/D\right)},\nonumber\\
& \text{with}\;\; & \lim_{H\to 0}\, \frac{H}{4D}\,\bar\rho \; =  \frac1{b-a} = \rho(x,t\to\infty)\;\;\text{as required.}
\eeqa

Whilst the weak noise approximation inherent in the saddle-point evaluation of path integrals is exact in the $V=0$ case, the preceding analysis reveals an issue at its core: the tension between the weak-noise and long-time limits, which do not commute \cite{kikuchi2020}. Indeed, the Laplace parameter $s = H/4D$, and care is required with the standard Laplace transform notion of ``small $s\leftrightarrow$ large $t$'' when $D$ is small. A path involving multiple circuits of the interval will have a very large $W$, and one might be tempted to neglect it. However, it is required for the correct evaluation of the long-time limit -- no matter how improbable a path may be made by the weakness of the noise, at sufficiently large times it will contribute. A partial explanation lies in the time-domain action, eq.\eqref{eq:legendre}: the term $-Ht$ acts to compensate a large-$H$ $W$ value (which $\sim 2\sqrt H$ at large $H$), increasingly so when $t$ becomes large as well. This suggests that, in the $H$-domain, the important criteria for inclusion is that the path must satisfy the Euler-Lagrange equations for the stochastic action, rather than be a small-action path \emph{per se}. 

The standard way to approach diffusion equations with finite boundary conditions is to use the method of images, and there is a one-to-one correspondence between the bouncing paths considered here and image terms. The paths which bounce off $x=a$ and $b$ can be identified as images located just beyond the boundaries, and the series of higher-order paths are the images of the images. The path integral approach offers considerable scope for generalization to nonzero potentials, where the method of images does not obviously apply \cite{chou2007}.

\section{Example: the harmonic oscillator}

In this section, we apply the results above to the harmonic oscillator, with potential $V = \frac1{2}\alpha x^2$ and infinite boundary conditions $\rho(x,t)\to 0$ as $x\to\pm\infty$. We take the starting point $x_0 =0$ for simplicity. The time-domain solution is the well-known Ornstein-Uhlenbeck (OU) density, 
\eq
\rho_{\rm OU}(x,t) = \sqrt{\frac{\alpha}{2\pi D(1-\ex^{-2\alpha t})}}\exp\left(\frac{-\alpha x^2}{2D(1-\ex^{-2\alpha t})}\right).\label{eq:OU}
\eeq Applying eq.\eqref{eq:rhobar}, the integrals can be performed explicitly, giving
\eq
W(x;H)=\alpha x^2+\frac1{\alpha}\left(\alpha x\,\sqrt{H+\alpha^2 x^2} - H\log\left(
\frac{-\alpha x + \sqrt{H+\alpha^2 x^2}}{\sqrt H}\right)\right),
\eeq and hence
\eqa
\bar\rho(x;H) & = & \calJ\exp\left(-\frac{W(x,H)}{4D}\right)\left(H\left(H+\alpha^2 x^2\right)\right)^{-1/4}\nonumber\\
\rho(x,t) & = & \frac1{2\pi i}\int_{\rm Br}\calJ\exp\left(+\frac{Ht}{4D}\right)\exp\left(-\frac{W(x,H)}{4D}\right)\left(H\left(H+\alpha^2 x^2\right)\right)^{-1/4}
\frac{\dee H}{4D}
\eeqa for $\bar\rho$ and $\rho$, where the Laplace inversion is performed along the Bromwich contour. The exponent is not quadratic in $H$, so the method of steepest descents will not give the integral exactly. Moreover, the expression $\bar\rho$ does not solve the transformed problem exactly either (Appendix B), so even if we could perform the inversion integral exactly, we would not expect to recover the OU solution. Proceeding optimistically with steepest descents, the stationary point of the exponent is at $H_*$, defined by eq.\eqref{eq:H}  
\eq
t = \frac{\partial W}{\partial H}\bigg|_{H=H_*} =  \int_0^x\frac{\dee y}{\sqrt{H_* + \alpha^2y^2}} = \frac1{2\alpha}\log\left[ \frac{\alpha x+\sqrt{H_* +\alpha^2 x^2}}{-\alpha x+\sqrt{H_* +\alpha^2 x^2}}\right], 
\eeq which can be inverted to give
\eq
H_* =  \alpha^2x^2\,{\rm cosech}^2\alpha t\in \mathbb{R}^+.\label{eq:H*}
\eeq 
\eq
W(H_*) = \alpha x^2\left(1+{\rm coth}\,\alpha t\right) + \alpha^2 x^2\,t\,{\rm cosech}^2\alpha t
 = \alpha x^2\left(1+{\rm coth}\,\alpha t\right)
 +H_*t,
 \eeq so 
 \eq
S(x,t) = \alpha x^2\left(1+{\rm coth}\,\alpha t\right) = 2\alpha x^2/(1-\ex^{-2\alpha t}).
 \eeq The contour is then the line $H=H_*+iy$, $y\in \mathbb{R}$, and steepest descents gives
\eq
\rho(x,t) = \frac{\calJ\exp(-S/4D )}{\left(H_*\left(H_*+\alpha^2 x^2\right)\right)^{1/4}}\sqrt{\frac1{8\pi D|W''(H_*)|}}. 
\eeq The integral for $W''$ (see eq.\eqref{eq:detL}) can also be done explicitly (see Appendix 1), and after some simple manipulations and the use of $\calJ = \exp(\alpha t/2)$\footnote{Note that within the steepest descents approximation, since $\calJ$ is independent of the small noise $D$, it is not involved in the integral, and can be thought of simply as $\calJ(t(H_*))$ in the Laplace domain, as can the 1/4 power prefactor term.}, this reduces exactly to the OU form eq.\eqref{eq:OU}. So, an approximate inversion of an approximate solution in the Laplace domain results in the exact time-domain solution. Of course, it would be too much to hope that the exactness would extend to more complicated potentials, but this does illustrate the surprising efficacy of the semiclassical approximation.

\section{Conclusions}

In this paper we have revisited the path integral formulation of stochastic processes. A closer look at the correspondence between the dominant stochastic paths and conservative Hamiltonian mechanics in an effective potential allows the conserved ``energy'' $H$ (which has dimensions of power) to be identified as the Laplace parameter of the diffusive dynamics. $H\to 0,\,t\to\infty$ is interpreted as the equilibrium limit. $H=0$ paths satisfy $\dot x= \mp V'$, and correspond to either noiseless relaxation to a local potential minimum, or ``optimal'' (probability-maximsing) hill-climbing respectively. As might be expected from comparing the Schr\"odinger equation with the Smoluchowski equation, the Laplace transform of a diffusive process plays an analogous role to the Fourier transform in quantum mechanics. The latter is governed by oscillatory wave-like dynamics, whereas the stochastic case has decaying relaxational modes. Working in the Laplace domain facilitates summing over trajectories involving multiple circuits of some interval, because the series of contributing terms to the transition probability density is geometric. Indeed, including an infinite series of such turning paths is essential for correctly evaluating the long-time, equilibrium limit. Finally, we discussed a curious error cancellation in the harmonic oscillator, where the approximate Laplace inversion of an approximate solution in the Laplace domain led to the exact time domain solution. 

\section*{Acknowledgments}

SPF acknowledges many useful discussions with Profs A. Archer and A. McKane. This work was supported by the UK EPSRC, grant number EP/R005974/1.

\section*{References}
\bibliography{paper}

\appendix
\section{Fluctuation determinant }

In this Appendix, we derive the explicit form of det$\,\hat L$, and discuss its long-time limit in the case of the harmonic oscillator. A convenient form for the determinant is given by (see e.g. \cite{schulman,kleinert})
\eq
{\rm det}\,\hat L = \left|\text{det}\,\frac{\partial x_{*i}(t)}{\partial\dot x_{*j}(0)}\right| = \left|\frac{\partial x_*(t)}{\partial\dot x_*(0)}\right|\;\;\text{in 1D}.
\eeq It is the derivative of the final position with respect to the initial momentum, and as such can be thought of as a ``density of paths''. On an extremal path $x_*$, $H = \dot x_*^2 - V'^2$ is 
constant. Dropping the ${}_*$ subscript and labelling $x(0)=x_0,x(t)=x_1$, evaluating $H$ at $t=0$ gives 
\eq
\frac{\partial}{\partial \dot x_0} = \frac{\partial H}{\partial
\dot x_0} \frac{\partial}{\partial H} = 
2\dot x_0  \frac{\partial}{\partial H} = 2\sqrt{H+V'(x_0)^2} 
 \frac{\partial}{\partial H};
\eeq (initial momentum and position are independent). Varying $H\to H+\delta H$, 
while holding $t$ constant $(\delta t = 0)$,
\eqa
t+\delta t & = & \int_{x_0}^{x_1+\delta x_1}\frac{\dee x}{\sqrt{H+\delta H + V'^2}}\nonumber\\
& = & \int_{x_0}^{x_1+\delta x_1}\frac{\dee x}{\sqrt{H + V'^2}} 
\left(1 -\frac1{2}\frac{\delta H}{H+V'^2} + 
...\right)\;\;\;\nonumber\\
& \approx & t + \delta x_1 \frac1{\sqrt{H + V'(x_1)^2}} 
-\frac{\delta H}{2}\int_{x_0}^{x_1}\frac{\dee x}{(H+V'^2)^{3/2}}
,
\eeqa so to keep $t$ constant, we require 
\eq
\frac{\delta x_1}{\delta H} = \sqrt{H + V'(x_1)^2} \cdot 
\frac1{2}\int_{x_0}^{x_1}\frac{\dee x}{(H+V'^2)^{3/2}}, 
\eeq and hence 
\eq
\frac{\partial x_1}{\partial \dot x_0} = 
\sqrt{H+V'(x_0)^2} \sqrt{H + V'(x_1)^2} 
\int_{x_0}^{x_1}\frac{\dee x}{(H+V'^2)^{3/2}}. 
\eeq If $V$ has a critical point on the path, then the integral in the above diverges as $H\to 0$. This is a manifestation of the well-known zero mode problem encountered when taking the long-time limit of fluctuation determinants (see e.g \cite{tarlie}). This is cancelled by the Jacobian, at least the in case of the harmonic oscillator. Taking $V=\frac1{2}\alpha x^2$ and $x_0=0$ for simplicity, 
\eq
\int_{0}^{x_1}\frac{\dee x}{(H+V'^2)^{3/2}}
=\frac{x_1}{H\sqrt{H+\alpha^2 x_1^2}}\to\infty\;\text{as}\; H\to 0,
\eeq and eq.\eqref{eq:H*} gives
\eq
H =  \alpha^2x_1^2\,{\rm cosech}^2\alpha t;\;\;\sqrt{H+\alpha^2 x_1^2} = \alpha x_1\,{\rm coth}\,\alpha t.
\eeq This results in
\eq
\frac{\partial x_1}{\partial \dot x_0} = \frac{\sinh\,\alpha t}{\alpha};\;\; \rho = \sqrt{\frac{\alpha}{\sinh\,\alpha t}}\,\calJ\exp{-\left(S/4D\right)}. 
\eeq Were it not for $\calJ$, this would $\to 0$ as $t\to\infty$, but the initially somewhat worrying-looking $\calJ = \exp\left( +\frac1{2}\alpha t\right)$ means the correctly-normalised OU density eq.\eqref{eq:OU} is recovered, with the long-time limit
\eq
\rho_{\rm eq}(x)=\sqrt{\frac{\alpha}{2\pi D}}\exp\left(-\frac{\alpha x^2}{2D}\right).
\eeq

\section{WKB method}

The Laplace-transformed Smoluchowksi equation is 
\eq
s\bar\rho(x;s) - \delta(x-x_0) = \frac{\dee}{\dee x}\left(\bar\rho V'+D\frac{\dee \bar\rho}{\dee x}\right). 
\eeq Inserting the WKB ans\"atz
\eq
\bar\rho = \exp\left( -\frac1{D}\sum_{n=0}^{\infty} D^nf_n(x)\right) = A(x)\exp\left(-\frac{W(x)}{4D}\right) + \text{h.o.t.}   ,
\eeq and keeping the first two terms only leads to  
\eq
As \approx V'A'-\frac{AV'W'}{4D}+AV''+DA''-\frac{AW''}{4}
-\frac{A'W'}{2}+\frac{AW'^2}{16D} .
\eeq Identifying $s=H/4D$, and thus including it in the dominant balance, immediately gives
\eq
W' = 2V'\pm 2\sqrt{H+V'^2};\; W = 2V\pm2\int\sqrt{H+V'^2}\,\dee x
\eeq exactly as with the path integral, with $\pm$ corresponding to right and left-moving paths respectively. If we had excluded the Laplace parameter from the dominant balance, we would have found $W=4V$, which is only correct in the long-time, time-independent equilibrium limit. The $\calO (1)$ terms give
\eqa
0 & = &  A'\left(V'-\frac{W'}{2}\right) + A\left(V''-\frac{W''}{4}\right);   \nonumber\\
\log A & = & -\frac1{4}\log\left(H+V'^2\right) \pm\frac1{2}\int^x V''\frac{\dee x}{\sqrt{H+V'^2}}\nonumber\\
A & = & \left(H+V'^2\right)^{-1/4} \exp\left(+\frac1{2}\int_0^tV''(x(t))\dee t\right),
\eeqa where the last line follows using $\dot x = \pm\sqrt{H+V'^2}$. (Within the steepest descents approximation, since $\calJ$ is independent of the small noise $D$, it is not involved in the integral, and can be thought of simply as $\calJ(t(H_*))$ in the Laplace domain). So, the 1/4 power prefactor term and the Jacobian emerge from a truncated WKB expansion, as expected by analogy with the quantum-mechanical case. The path integral formula gives the correct normalizing factor, and prescription for left and right-moving paths, whereas the full WKB solution would have to be constructed by explicitly matching solutions,and imposing the appropriate jump condition on $\bar\rho'$ at $x=x_0$. However, even for the harmonic potential $V=\alpha x^2/2$, this is not exact at this order: the term $DA''$ is neglected. By approximately inverting the transform using steepest descents, the exact time domain (OU) solution is recovered.

\end{document}